\documentclass[preprint,preprintnumbers,amsmath,amssymb,titlepage]{revtex4-2}
\usepackage{graphicx}
\usepackage{comment}
\usepackage[normalem]{ulem}
\usepackage{bm}
\usepackage[hidelinks]{hyperref}
\usepackage{color}
\usepackage[symbol,bottom]{footmisc}

\usepackage{dcolumn}
\usepackage[right]{lineno}

\newcommand{\tc}{$T_{\rm c}$}

\begin{document}

\title{
Higher-order epitaxy: A pathway to suppressing structural instability and emergent superconductivity
}




\author{Yuki Sato$^{1}$}\email{\ yuki.sato.yj@riken.jp}
\author{Soma Nagahama$^{2}$}
\author{Shunsuke Kitou$^{3}$}
\author{Hajime Sagayama$^{4}$}
\author{Ilya Belopolski$^{1}$}
\author{Ryutaro Yoshimi$^{1, 3}$}
\author{Minoru Kawamura$^{1}$}
\author{Atsushi Tsukazaki$^{2, 5}$}
\author{Naoya Kanazawa$^{6}$}
\author{Takuya Nomoto$^{7}$}
\author{Ryotaro Arita$^{1, 8}$}
\author{Taka-hisa Arima$^{1, 3}$}
\author{Masashi Kawasaki$^{1, 2}$}
\author{Yoshinori Tokura$^{1, 2, 9}$}\email{\ tokura@riken.jp}

\affiliation{$^{1}$RIKEN Center for Emergent Matter Science (CEMS), Wako 351-0198, Japan} 
\affiliation{$^{2}$Department of Applied Physics and Quantum-phase Electronics Center (QPEC), University of Tokyo, Tokyo 113-8656, Japan}
\affiliation{$^{3}$Department of Advanced Materials Science, University of Tokyo, Kashiwa 277-8561, Japan}
\affiliation{$^{4}$Institute of Materials Structure Science, High Energy Accelerator Research Organization, Tsukuba 305-0801, Japan}
\affiliation{$^{5}$Institute for Materials Research (IMR), Tohoku University, Sendai 980-8577, Japan}
\affiliation{$^{6}$Institute of Industrial Science, University of Tokyo, Tokyo 153-8505, Japan}
\affiliation{$^{7}$Department of Physics, Tokyo Metropolitan University, Hachioji, Tokyo 192-0397, Japan}
\affiliation{$^{8}$Department of Physics, University of Tokyo, Tokyo 113-0033, Japan}
\affiliation{$^{9}$Tokyo College, University of Tokyo, Tokyo 113-8656, Japan}

\date{\today}
\maketitle

\begin{center}
    \textbf{Abstract}
\end{center}

Molecular beam epitaxy enables the growth of thin film materials with novel properties and functionalities. Typically, the lattice constants of films and substrates are designed to match to minimise disorders and strains. However, significant lattice mismatches can result in higher-order epitaxy, where commensurate growth occurs with a period defined by integer multiples of the lattice constants. Despite its potential, higher-order epitaxy is rarely used to enhance material properties or induce emergent phenomena. Here, we report single-crystalline FeTe films grown via 6:5 commensurate higher-order epitaxy on CdTe(001) substrates. Scanning transmission electron microscopy reveals self-organised periodic interstitials near the interface, arising from higher-order lattice matching. Synchrotron x-ray diffraction shows that the tetragonal-to-monoclinic structural transition in bulk FeTe is strongly suppressed. Remarkably, these films exhibit substrate-selective two-dimensional superconductivity, likely due to suppressed monoclinic distortion. These findings demonstrate the potential of higher-order epitaxy as a tool to control materials and inducing emergent phenomena.


\newpage

Engineering materials in thin film form not only enhances our understanding of quantum phenomena but also provides knobs to tune material properties. Among various deposition techniques, molecular beam epitaxy (MBE)--- a method where films are grown layer by layer with lattice matching to a substrate--- is one of the most powerful tools for fabricating a wide variety of thin films and heterostructures. It offers exceptional control over structure and composition, enabling novel properties and functionalities that go beyond those of bulk crystals \cite{arthur2002molecular, goldman1999cuprate, pfeiffer2003role, wang2012interface, tokura2019magnetic, nunn2021review, naritsuka2021controlling}. In materials engineering, achieving high-quality materials with minimal disorders, such as defects, misorientation and dislocations, is crucial since quantum transport phenomena often favour clean systems with long scattering times \cite{pfeiffer2003role}. In contrast, it is widely known that significant lattice mismatches between a film and a substrate can sometimes result in higher-order epitaxy, also referred to as domain-matching epitaxy, where commensurate growth occurs with a long period defined by integer multiples of the lattice constants \cite{stoyanov1986theory, narayan2003domain, springholz2001nanoscale, estandia2020domain}. Although such growth modes are typically avoided as they introduce unnecessary disorder into crystals, it has recently been suggested that spatially modulated strain, caused by mismatches, can be used to control the topological properties of materials \cite{tang2014strain}. Despite its potential, there are few examples of higher-order epitaxy being utilised to enhance materials properties or induce novel quantum phenomena \cite{zeljkovic2015strain}.

Iron chalcogenides are one of the most attractive materials platforms for studying the interplay between thin films and substrates, as exemplified by the significant enhancement of the superconducting transition temperature $T_{\rm c}$ in monolayer FeSe grown on SrTiO$_3$ (STO), from a bulk value of 10 K to above 50 K in the monolayer form \cite{wang2012interface}. The Fe(Se,Te) (FST) system has attracted significant interest as a candidate topological superconductor driven by strong spin-orbit coupling from Te, where a Majorana zero mode is expected to emerge at the core of vortex \cite{wang2015topological, zhang2018observation, wang2018evidence}. It is also utilised for a building block for an artificial topological superconductor, where FST is proximitised with topological insulators \cite{Sato2024Molecular, Nagahama2025Control}. FeTe is the other end compound of the FST series and is a non-superconducting antiferromagnetic (AFM) metal. It has a tetragonal $P$4/$nmm$ crystalline structure consisting of van der Waals-stacked iron-chalcogenide square lattices at room temperature (Fig. \ref{fig:Xtal}a). FeTe crystals with minimal excess iron amount undergo a transition to the monoclinic $P$2$_1$/$m$ phase at the structural transition temperature $T_{\mathrm s} \sim$ 70 K, below which in-plane anisotropy develops ($a/b > 1$) and the $c$-axis tilts slightly towards the $a$-axis ($\beta < 90^{\circ}$) (Fig. \ref{fig:Xtal}b) \cite{koz2013low}. This monoclinic lattice distortion is accompanied by the emergence of a double-stripe AFM order with a magnetic vector $Q$ = ($\pi$, 0) in the reciprocal space of the tetragonal lattice with two iron atoms per unit cell \cite{neutron2009bao}. Neutron scattering experiments reveal that superconductivity emerges when the ($\pi$, 0) magnetic correlations are suppressed by partly substituting Se for Te, suggesting that the AFM order is detrimental to superconductivity \cite{liu2010pi}.

In this paper, we report growth of single-crystalline higher-order epitaxial superconducting FeTe films using the MBE technique grown on cubic zinc-blende CdTe(001) substrates (Fig. \ref{fig:Xtal}c), (See Methods and \cite{Sato2024Molecular} for more details about sample fabrication). For comparison, we also grew FeTe films on cubic perovskite STO(001) substrates (Fig. \ref{fig:Xtal}d). FeTe/CdTe films appear to be single-crystalline with the crystalline orientation of FeTe highly aligned to that of CdTe. Figure \ref{fig:Xtal}e shows an XRD $\theta$-$2\theta$ profile for a typical FeTe/CdTe film. We observe sharp peaks attributed to FeTe (0 0 $n$) or CdTe (0 0 2$m$), where $n$ and $m$ are integers, indicating that the FeTe (0 0 1) plane is aligned parallel to the substrate normal, i.e., FeTe [001] $\parallel$ CdTe [001]. The high crystalline quality of the film is further supported by a rocking curve for the FeTe (0 0 1) reflection. As shown in Fig. \ref{fig:Xtal}f, the rocking curve for FeTe/CdTe exhibits a sharp signal with a small full-width at half maximum (FWHM) of 0.6$^{\circ}$, whereas that for FeTe/STO shows a much broader value of 1.8$^{\circ}$. Furthermore, XRD azimuthal angle $\varphi$-scans for the asymmetric Bragg reflections of FeTe (1 0 4) and CdTe (1 1 5) show clear 4-fold oscillations of FeTe (1 0 4), following the same trend as CdTe (1 1 5) (Fig. \ref{fig:Xtal}g). The result is consistent with the tetragonal crystal structure and indicates a high lateral orientation of the FeTe film, where FeTe [100] $\parallel$ CdTe [110]. This lateral orientation is further confirmed by atomic force microscopy measurements, which reveal well-defined square-patterned steps and terraces with edges aligned to CdTe $\langle110\rangle$ (Fig. \ref{fig:Xtal}h). Remarkably, FeTe/CdTe films exhibit superconductivity with an onset critical temperature $T_{\rm c} \sim$ 12 K, whereas FeTe/STO films grown under the exactly same conditions do not show any superconducting onset down to 2 K (Fig. \ref{fig:Xtal}i). 
The zero resistivity in FeTe/CdTe, which was not reported in our previous works \cite{yasuda2019nonreciprocal, Sato2024Molecular}, is achieved by further optimizing growth conditions, such as growth rate and Te-annealing temperature and time. While it has been reported that superconductivity in FeTe can be induced by oxygen incorporation, this typically requires aggressive methods such as exposure to atmosphere for weeks or annealing under oxygen flow \cite{telesca2012impact, zhang2013superconductivity}. To rule out this possibility, we simultaneously synthesised, in-situ Te-capped, and measured those two films within a day. The results suggest that the substrate-selective superconductivity is an intrinsic property of FeTe/CdTe. These observations motivate a detailed investigation into the crystal structure of FeTe/CdTe.


It is worth noting that the in-plane lattice mismatch between FeTe ($a_{\mathrm{FeTe[100]}}$ = 0.383 nm) and CdTe ($a_{\mathrm{CdTe[110]}}$ = 0.458 nm) is as large as +20\%, whereas that between FeTe and STO ($a_{\mathrm{STO[100]}}$ = 0.390 nm) is relatively small at $-2\%$. Given this, it is somewhat surprising that FeTe/CdTe films exhibit apparently higher crystalline quality than FeTe/STO films (See Extended Data Fig. \ref{fig:FTSTO} for further characterisation of FeTe/STO). This counterintuitive result can be attributed to higher-order epitaxy, characterised by a 6:5 commensuration between FeTe and CdTe, as we show in the following. Figure \ref{fig:STEM}a displays a cross-sectional high-angle annular dark-field scanning transmission electron microscopy (HAADF-STEM) image of the FeTe/CdTe interface viewed from FeTe [100] $\parallel$ CdTe [110] at room temperature. The interface between FeTe and CdTe is clearly resolved with atomic resolution, as indicated by the overlaid cartoons. Figure \ref{fig:STEM}b presents line profiles along cut A (the bottommost layer of FeTe) and B (the topmost layer of CdTe). While the two profiles differ in modulation period due to the substantial lattice mismatch, there are distinct "nearest sites" where two atoms closely approach each other. These sites recur every 6$a_{\mathrm {FeTe[100]}}$ or 5$a_{\mathrm {CdTe[110]}}$, as highlighted by the grey-shaded regions in Fig. \ref{fig:STEM}b. This higher-order lattice matching results in a considerably small residual mismatch, calculated as (6$a_{\mathrm {FeTe[100]}}-5a_{\mathrm {CdTe[110]}}$)/6$a_{\mathrm {FeTe[100]}} \sim$ 0.34\%. These results confirm the successful realisation of higher-order epitaxy in FeTe/CdTe films.

The higher-order epitaxy appears to be stabilised by the introduction of self-organised interstitials and periodic displacement only near the interface. As highlighted by the red arrows in Fig. \ref{fig:STEM}b, a pair of additional peaks is observed adjacent to the nearest sites. These interstitials, likely intercalated to reduce interfacial energy associated with the higher-order epitaxy, do not belong to the crystal structures of either FeTe or CdTe, breaking the translational symmetry of FeTe. Such interfacial interstitials have not been observed in other van der Waals epitaxial systems such as FeSe/STO and FeSe/MgO \cite{Li2016Atomically, Peng2020picoscale, obata2021chemical}. Typically, higher-order epitaxial films relax the strain through periodic or random dislocations, characterised by Burgers vectors oriented in the in-plane direction \cite{springholz2001nanoscale, terai2002plane, ren2022rotation}. However, in the present system, no apparent dislocation is observed at the interface, and both FeTe and CdTe retain their original crystal structure except for the interstitials. Moreover, we found a slight displacement of atoms at certain sites, marked as 2 and 4 in Fig. \ref{fig:STEM}b, from the equilibrium position. This displacement can be viewed as a strain that is not uniform, but spatially modulates.

We examine a HAADF-STEM image of an FeTe/CdTe interface over a wider observation frame (Fig. \ref{fig:STEM}c). Interestingly, the interstitials at the FeTe/CdTe interface extend slightly along FeTe [001] into a few unit cells, forming discernible \textit{nodes} that periodically recur along the FeTe [010] direction, spanning the entire observation window. Figure \ref{fig:STEM}d displays the Fourier transform image derived from the HAADF-STEM image near the interface, labelled as "FFT1" in Fig. \ref{fig:STEM}c. In addition to the Bragg peaks corresponding to FeTe and CdTe, we observe additional peaks associated with the modulation from higher-order epitaxy. To highlight this, Fig. \ref{fig:STEM}e shows a profile integrated along the $Q_c$ direction as a function of $Q_a$. Here the horizontal axis is normalised by the reciprocal lattice vector of FeTe $Q_{\rm{FeTe}}$. Apart from the primary Bragg peaks, a distinct peak appears at $Q_{a}/Q_{\rm{FeTe}} = 1/6$, consistent with the modulation period of $6a_{\mathrm {FeTe}}$. This peak indicates that the interstitials extend coherently over the interface and create intrinsic structural modulation rather than extrinsic effects, such as moir\'{e} interference patterns along the electron beam direction. In addition to the primary modulation mode at $Q_{a}/Q_{\rm{FeTe}} = 1/6$, related modes at $Q_{a}/Q_{\rm{FeTe}} = 2/6$, 3/6, and 4/6 are faintly resolved. A mode at $Q_{a}/Q_{\rm{FeTe}} = 1/12$ appears with an intensity somewhat comparable to the primary 1/6 mode, hinting at the possible involvement of a secondary modulation with a period of $12a_{\mathrm {FeTe}}$ in stabilising the interface. When the FFT is performed on the image taken 3 UCs away from the interface (FFT2), the peak at $Q_{a}/Q_{\rm{FeTe}} = 1/6$ is not discernible. This suggests that the higher-order epitaxy occurs only near the interface, transitioning to conventional van der Waals epitaxy at a greater distance. Figure \ref{fig:STEM}f schematically illustrates the FeTe/CdTe interface from the top (FeTe [001]), where the interstitial pairs form superlattice along the interface, resulting in a characteristic $6\times6$ modulation. This modulation is also confirmed by Reflection High Energy Electron Diffraction (RHEED), which reveals additional pronounced peaks corresponding to the superlattice only when a few UCs of FeTe are deposited (Extended Data Fig. \ref{fig:RHEED}). Identifying the chemical nature of the interstitials, however, remains challenging, as their close spacing with other elements exceeds the resolution limits of our current experiments.

As higher-order epitaxy is a characteristic of FeTe/CdTe films, it is tempting to attribute the substrate-selective superconductivity to some consequences resulting from this epitaxy. To access a potential link between the superconductivity and crystal structure at low temperatures, we performed temperature ($T$)-dependent synchrotron XRD experiments on FeTe films. First, as a reference, we measured a thick FeTe/STO film with a thickness $t$ of 1000 nm, denoted as FeTe1000/STO (Fig. \ref{fig:FTCT}a). At $T$ = 100 K, (0 0 7) Bragg reflection appears as a single peak in the $\theta$-2$\theta$ profile. As $T$ decreases, the intensity of this peak steadily diminishes, while another peak begins to emerge at lower angles (2$\theta\sim$ 70.7$^{\circ}$) below $T_{\rm s} \sim$ 60 K. At a base temperature of 7 K, the higher-angle peak disappears completely, leaving a single peak at the lower angle. This discrete shift in the 2$\theta$ peak position is a consequence of the tetragonal-to-monoclinic structural transition, as reported in bulk crystals \cite{neutron2009bao, XRD2011Xiao} (See also Methods for reciprocal space mappings).

Next, we performed $\theta$-$2\theta$ XRD scans for thick and thin FeTe/CdTe films with $t$ = 1000 nm (FeTe1000/CdTe) and 40 nm (FeTe40/CdTe), as shown in Figs. \ref{fig:FTCT}b and \ref{fig:FTCT}c. Hereafter, FeTe$t$/CdTe denotes a $t$-nm-thick FeTe film grown on CdTe(100) substrate. In the thick FeTe1000/CdTe film, similar to the FeTe1000/STO film, the lower-angle peak associated with the monoclinic phase develops below $T_{\rm s}$. Interestingly, in FeTe1000/CdTe, the higher-angle peak associated with the tetragonal phase remains prominent even below $T_{\rm s}$, and the $\theta$-2$\theta$ scan reveals a distinct double-peak structure at low temperatures. This double-peak structure indicates the coexistence of tetragonal and monoclinic phases in FeTe1000/CdTe. Notably, the tetragonal phase exhibits a stronger intensity compared to the monoclinic phase, suggesting that, contrary to bulk crystals, the tetragonal phase dominates in FeTe1000/CdTe at low temperatures. The contrasting behaviour observed in FeTe1000/STO underscores the critical role of substrate-induced effect, such as cramping, in suppressing monoclinic distortion in FeTe1000/CdTe. Furthermore, in the thin FeTe40/CdTe film, the tetragonal phase is even more dominant, as evidenced by the $\theta$-2$\theta$ scans, which show a single major peak with slight asymmetry developing below $T_{\rm s}$. This asymmetry is attributed to a small portion exhibiting monoclinic distortion. The double Gaussian fitting at each temperature successfully separates the two components in the data.

In Fig. \ref{fig:FTCT}d, we summarise the out-of-plane lattice constant $c$, estimated using single or double Gaussian fittings, as a function of $T$, alongside values for bulk crystals reported in the literature \cite{neutron2009bao, XRD2011Xiao}. At 300 K, the $c$-values of the films are close to those of the bulk crystals, suggesting that strain from the substrates is minimal. Upon cooling, $c$ of the films decreases, with the largest slope of $dc/dT$ observed in FeTe40/CdTe, indicating a substantial strain effect from CdTe. The variation in $dc/dT$ can be attributed to the difference in thermal expansion between FeTe and the substrates. The smaller lateral thermal expansion of CdTe leads to an effective in-plane tensile strain in FeTe, which in turn enhances the out-of-plane thermal shrinkage in the films (Extended Data Figs. \ref{fig:thermal}a-b). The Poisson ratios $\nu$ of FeTe1000/CdTe and FeTe40/CdTe, evaluated from the $dc/dT$ slope, correspond to 0.4 and 0.5, respectively, showing an enhancement compared to the bulk value of 0.2 \cite{chandra2010elastic} (See Methods for more details). This enhanced Poisson ratio suggests a considerably strong coupling between FeTe and CdTe, which plays a crucial role in achieving the cramping effect that prevents the structural transition. Below $T_s$, the $c$-value of FeTe/CdTe films exhibits a clear split, with the tetragonal branch subsisting while a new branch appears in the monoclinic phase, as highlighted in the grey area of Fig. \ref{fig:FTCT}d. In this phase, $c$ for all the films appears to approach the bulk value at low temperatures, suggesting complete lattice relaxation free from the strain effects.

To investigate the impact of substrate effects on the AFM order in FeTe, we performed first-principles calculations, where the in-plane lattice constant was fixed to mimic the substrate constraint (See Methods for more details). With decreasing $-\Delta c/c$, which corresponds to the uniaxial compressive strain along the $c$-axis, the Fe moments ($\mu$) monotonously decreases, while the energy difference between the AFM and paramagnetic ground states, $E_{\rm AFM} - E_{\rm PM}$, increases (Fig. \ref{fig:FTCT}e). This is likely because the compression strain along the $c$-axis enhances hopping via Te atoms, reducing electron correlations and making the AFM ordering less favourable. Indeed, in thinner FeTe/CdTe films, anomalies in magnetic susceptibility and the Hall coefficient at $T_{\rm s}$, associated with the AFM ordering, become less pronounced \cite{Sato2024Molecular}. These results suggest that the AFM ordering is considerably suppressed under the strong compressive strain along the $c$-axis.

It is informative to estimate the volume fraction of the monoclinic phase, $V_M/(V_T + V_M)$, where $V_M$ and $V_T$ represent the volumes of the monoclinic and tetragonal phases, respectively, derived from the integrated intensity of each Gaussian component in the $\theta$-2$\theta$ profiles. Figure \ref{fig:FTCT}f displays the $T$-dependence of $V_M/(V_T + V_M)$. For FeTe1000/STO, the monoclinic phase sharply develops below $T_{\rm s}$ and reaches nearly 100 \% within experimental detection limits below $T \sim$ 30 K. In contrast, the monoclinic phase is significantly suppressed in FeTe/CdTe films, with $V_M/(V_T + V_M)$ reaching about 40 \% for FeTe1000/CdTe and as low as 15 \% for FeTe40/CdTe. The $V_M/(V_T + V_M)$ value remains around 40 \% even for an FeTe200/CdTe film, suggesting that substantial suppression of the monoclinic phase requires films at least thinner than 200 nm (Extended. Data Fig. \ref{fig:200nm}). These results suggest that the suppression of the monoclinic phase is due to the substrate cramping effect, which extends remarkably far into the FeTe film, influencing regions up to 600 nm from the interface in FeTe1000/CdTe, as illustrated in the inset of Fig. \ref{fig:FTCT}f. This is somewhat unexpected, as the higher-order epitaxy occurs only within a few UCs near the interface, yet it appears to stabilise the crystal structure over a remarkably large length. This strong coupling between the substrate and the film likely plays a crucial role in the substrate-selective superconductivity, as the strong out-of-plane compressive strain and the suppression of the monoclinic phase weaken the concomitant AFM order, which is known to compete with superconductivity \cite{liu2010pi}.

We next investigated the superconducting properties of FeTe/CdTe films and found that they exhibit distinct two-dimensional characteristics. Figure \ref{fig:SC}a shows the $T$-dependence of resistivity for $t$ = 10, 20, 40 and 80 nm, where superconductivity is observed in all the films. This observation is consistent with our scenario that the superconductivity can emerge in films with suppressed tetragonal-to-monoclinic transition. Interestingly, the zero-resistivity critical temperature $T_{\rm c0}$ decreases with decreasing $t$, whereas the onset temperature $T_{\rm c}$ around 12 K is nearly independent of $t$. This nearly $t$-independent onset $T_{\rm c}$ suggests that the energy scale for superconducting gap formation is similar across all films. The broad superconducting transitions observed in thinner films are attributed to the Berezinskii-Kosterlitz-Thouless (BKT) transition, where strong fluctuations due to the low dimensionality prevent long-range order at finite temperatures. As shown in Fig. \ref{fig:SC}b and Extended Data Fig. \ref{fig:2DSC}a, the temperature dependence of resistivity for the $t$ = 10, 20 and 40 nm films is fitted by the Halperin-Nelson (HN) formula:
\begin{equation}
    R = R_0\exp{\left[ -2b\left(\frac{T_{0} - T}{T - T_{\rm BKT}}\right)^{1/2}\right]},
\end{equation}
where $R_0$ and $b$ are material-dependent fitting parameters \cite{halperin1979resistive}. The fitting for the FeTe40/CdTe film yields $T_{\rm BKT} = 5.8$ K. The current-voltage ($I$-$V$) characteristics for the FeTe40/CdTe film (Fig. \ref{fig:SC}c) show the power-law behaviour $V \propto I^{\alpha}$, as expected fro two-dimensional superconductivity. The inset of Fig. \ref{fig:SC}c displays the $T$ dependence of the exponent $\alpha$, which crosses $\alpha = 3$ at around $T_{\rm BKT}$ = 5.8 K, as expected from the BKT theory. 

Two-dimensional superconductivity in thin FeTe/CdTe films is further corroborated by the angle-dependence of the upper critical field $H_{\rm c2}$. Figure \ref{fig:SC}d displays the polar angle $\theta$ dependence of $H_{\rm c2}$, which is defined by the fields at which $R$ reduces to half of the normal-state value. The $H_{\rm c2}$ of the FeTe40/CdTe film exhibits a distinct kink anomaly at $\theta$ = 90$^{\circ}$, which is well captured by the two-dimensional Tinkham model, rather than the three-dimensional anisotropic mass model \cite{tinkham2004introduction}. In stark contrast, the $\theta$-dependence of $H_{\rm c2}$ for the FeTe80/CdTe film agrees well with the three-dimensional model (Extended Data Fig. \ref{fig:80nm}). At present, the origin of this crossover in superconducting dimensionality between $t$ = 40 and 80 nm remains unclear. One possible explanation is that the interface hosts two-dimensional superconductivity, whereas the bulk of the FeTe film exhibits three-dimensional superconductivity whose $T_{c0}$ exceeds that of the interface for thicknesses above 80 nm. 
The observation that only thin FeTe/CdTe films exhibit two-dimensional superconductivity underscores the crucial role of the substrate effect in the emergence of superconductivity in this system.

The superconductivity observed in FeTe/CdTe is reminiscent of that reported in heterostructure systems where FeTe is interfaced with other tellurides \cite{yao2025mystery}, such as Bi$_2$Te$_3$-based compounds \cite{yasuda2019nonreciprocal, he2014two, yao2021hybrid, yi2024interface}, MnBi$_2$Te$_4$ \cite{yuan2024coexistence}, MnTe \cite{yao2022superconducting}, and Fe$_3$GeTe$_2$ \cite{Taek2025universal}. Superconductivity has also been observed in FeTe thin films grown via pulsed laser deposition, where tensile strain--- similar to that observed in FeTe/CdTe--- is argued to be crucial for inducing superconductivity \cite{han2010superconductivity}. However, in all these systems, including heterostructures, structural investigation at low temperatures is lacking. To clarify the mechanism behind the emergence of superconductivity, future investigations are required to determine whether the tetragonal phase is preserved in these heterostructures or not. 

Finally, we comment on potential applications of the higher-order epitaxial FeTe films. 
One major obstacle in investigating Majorana fermions in the topological superconductor candidate FeSe$_{0.45}$Te$_{0.55}$ solid solution is the inevitable inhomogeneity associated with Se/Te clustering. ARPES studies have revealed that topological surface states tend to appear in a Te-rich regions \cite{li2021electronic}, highlighting the essential role of Te atoms in enhancing spin-orbit coupling. Inhomogeneity may also produce trivial conductance peaks at non-zero energies in scanning tunneling spectroscopy experiments \cite{machida2023searching}, which can mimic the zero-bias peaks expected from Majorana zero modes. By comparison, the superconducting properties of FeTe/CdTe appear similar to those of Te-rich FST films: it exhibits \tc\ $\sim$ 12 K and relatively isotropic upper critical fields (Fig. \ref{fig:SC}e and Extended Data Fig. \ref{fig:80nm}b), similarly reported in Te-rich FST thin films grown on CdTe \cite{Sato2024Non-Fermi}. Moreover, a recent ARPES study revealed topological surface states in an FeSe$_{0.07}$Te$_{0.93}$ thin film \cite{lin2025topological}, suggesting that topological states are preserved deep into the FeTe end of the series. These results suggest that the superconducting FeTe, with strong spin-orbit coupling yet free from compositional inhomogeneity, provides an ideal platform for investigating Majorana fermions.

Another intriguing direction of research involves the superlattice of interstitials in FeTe/CdTe, which breaks the translational symmetry of the original film. Moir\'e patterns have been observed in one-unit-cell FeTe grown on Bi$_2$Te$_3$ \cite{qin2021moire} and NbSe$_2$ \cite{deng2021moire}, where $8\times2$-type and stripe-like patterns, distinct from the present $6\times6$ pattern on CdTe, are respectively resolved. Such superlattices may generate a long-range moir\'e-like potential, analogous to effects seen in twisted bilayer graphene \cite{cao2018unconventional}, possibly modifying the electronic structure of the FeTe film near the interface. As higher-order epitaxy can be applied to a wide range of materials systems, our MBE-based technique may provide a more robust and reproducible approach than exfoliation-based approaches to realising superstructures with periodic strain and potential, opening new avenues for controlling the topological properties of materials and inducing quantum emergent phenomena \cite{tang2014strain, zeljkovic2015strain}.


\clearpage
\renewcommand\refname{References}
\bibliographystyle{naturemag}
\bibliography{FeTe}
\newpage

\textbf{METHODS}\\
\\
\textbf{Sample fabrication:}
All samples studied in this work were synthesised using the MBE method \cite{Sato2024Molecular}. Prior to the synthesis, CdTe(001) substrates were etched with bromine-methanol (0.1\% bromine) for 30 seconds, while STO substrates were annealed at 1000$^{\circ}$C for 10 hours under an O$_2$ flow. The films were synthesised at a substrate temperature of 240$^{\circ}$C in an ultrahigh vacuum chamber, with simultaneous beam fluxes of Fe and Te (plus Se for FST). The growth rate was approximately at 3 min/nm. After the growth, a Te-flux was continuously supplied for 30 minutes while samples cooled.\\

\textbf{X-ray diffraction}:
We performed XRD for room-temperature characterisation of samples using SmartLab (RIGAKU) with Cu K$\alpha$ x-ray. All other low-temperature XRD data was taken with synchrotron x-ray source (12 keV) at Photon Factory beam line 4C, High Energy Accelerator Research Organization (KEK) in Japan.\\

\textbf{Transport}:
We measure electrical transport properties with Quantum Design Physical Properties Measurement System (PPMS). We used indium soldering for electrical contact to films. The excitation current is fixed at 10 $\mu$A.\\

\textbf{Scanning transmission electron microscopy}:
The samples were cut into lamellae using focused ion beam technique and cleaned with Ar ion beam. We performed STEM measurements with acceralation voltage of 200 kV using JEM-ARM200F (JEOL) at Foundation for Promotion of Material Science and Technology of Japan (MST).\\

\textbf{HAADF-STEM study near an Fe(Se,Te)/CdTe interface}:
We performed HAADF-STEM experiments on an FeSe$_{0.1}$Te$_{0.9}$ (FST) thin film, where the higher-order matching is further reduced to (6$a_{\mathrm {FST[100]}}-5a_{\mathrm {CdTe[110]}}$)/6$a_{\mathrm {FST[100]}} \sim 1 \times$ 10$^{-3}$. Extended Data Fig. \ref{fig:STEM_FST}a displays a cross-sectional HAADF-STEM image of the FST/CdTe interface viewed from FST[100] $\parallel$ CdTe[110] at room temperature. Compared to the FeTe/CdTe interface (Fig. \ref{fig:STEM}a), the FST/CdTe interface exhibits a clearer atomic image, presumably due to the better higher-order matching. Extended Data Fig. \ref{fig:STEM_FST}b presents line profiles along cut A (the bottommost layer of FST) and B (the topmost layer of CdTe). Unlike FeTe/CdTe, interstitials are found only at the interface, adjacent to the nearest sites. Moreover, an oscillation mode is observed in the amplitude of the profile along cut A, which can be modeled as a sinusoidal wave with a period of $6a_{\mathrm{FST[100]}}$, expressed as $|\sin{(2\pi x/a_{\mathrm{FST[100]}})}|$ (dashed line in Extended Data Fig. \ref{fig:STEM_FST}b). This long-period oscillation suggests slight rippling in the FST layer, driven by interstitial pairs at the interface.\\

\textbf{Reciprocal space mappings}:
To check whether the tetragonal-to-monoclinic structural transition, as reported in bulk FeTe crystals, occurs in MBE-grown FeTe films, we performed reciprocal space mappings around the asymmetric FeTe (2 0 7) Bragg point for FeTe1000/STO. As shown in Fig. \ref{fig:RSM}a, at 80 K ($T > T_{\rm s}$), a clear single peak is observed around (2 0 7) reflection, as four ($\pm$2 0 7), (0 $\pm$2  7) reflections are degenerated in the tetragonal phase. On the other hand, at 7 K ($T < T_{\rm s}$), the (2 0 7) reflection splits into three peaks (Fig. \ref{fig:RSM}b). This splitting is consistent with the monoclinic phase, where in-plane anisotropy ($a/b \neq 1$) and a non-zero tilt angle ($\beta \neq \pi/2$) cause the ($\pm$2 0 7) reflections to split, while the (0 $\pm$2 7) reflections appear at the same reciprocal space vector. The intensity of the peaks is also consistent, where the doublet has doubled intensity than the two singlets. These results support that the peak splitting observed in the XRD $\theta$-2$\theta$ scans (for example, data at 50 K in Fig.\ref{fig:FTCT}a) are attributed to the monoclinic and tetragonal phases, respectively. Extracting the peak centres from Fig. \ref{fig:RSM}b, we obtain $a$ = 0.38403 nm, $b$ = 0.37966 nm, $c$ = 0.62529 nm, and $\beta$ = 88.647$^{\circ}$, which reasonably agree with those of bulk single crystals \cite{neutron2009bao}. Overall, the structural transition observed in the FeTe1000/STO is akin to that of bulk crystals, indicating no discernible strain from the STO substrate.\\

\textbf{Strain due to thermal expansion mismatch}:
A lattice mismatch between substrates and epitaxial films results in tensile or compression strain, which sometimes affects electronic properties of the films \cite{phan2017effects}. To assess strains in FeTe films, we assume that a CdTe substrate clamps the FeTe film so perfectly that the in-plane lattice constant 6$a_{\mathrm{FeTe}}$ follows the same temperature dependence as 5$a_{\mathrm{CdTe[110]}}$ (Extended Data Fig. \ref{fig:thermal}a). We restrict ourselves to quantify the strain only down to $T_{\mathrm s} \sim$ 70 K, because the crystal structure of bulk and film FeTe are different below $T_{\mathrm s}$. The temperature dependence of the in-plane strain $\varepsilon_a$ is calculated as $\varepsilon_a = (6a_{\mathrm{FeTe\hspace{0.5mm}bulk}} - 5a_{\mathrm{CdTe[110]}})/6a_{\mathrm{FeTe\hspace{0.5mm}bulk}}$ (Extended Data Fig. \ref{fig:thermal}b). At 300 K, $\varepsilon_a$ is estimated to be as small as +0.03$\%$, indicating good higher-order lattice matching results in a tiny tensile strain at room temperature. With decreasing $T$, $\varepsilon_a$ turns to be negative, and shows a considerable value of about $-0.2\%$ at 70 K. This in-plane compression strain leads to the out-of-plane expansion strain $\varepsilon_c$, which is expressed as $\varepsilon_c = 2\varepsilon_a/(1 - 1/\nu)$ \cite{Poisson_eq}, where $\nu$ is the Poisson ratio. Indeed, the steeper slope in the temperature dependence of $c$ for FeTe films indicates a significant out-of-plane expansion strain from substrates (Fig. \ref{fig:FTCT}d and Extended Data Fig. \ref{fig:200nm}b). $\nu$ for FeTe1000/CdTe, FeTe200/CdTe, and FeTe40/CdTe is estimated to be 0.5, 0.45, and 0.4, respectively, which is significantly larger than 0.21, which was obtained by a first-principles calculation for bulk FeTe \cite{chandra2010elastic}. Such a significant enhancement of the Poisson ratio in epitaxial films is also reported in graphene, where the $\nu$ value is typically about 0.19 but almost doubled to 0.36 when grown on Ni(111) substrates \cite{politano2015probing}. This enhancement in $\nu$ is claimed to be due to a commensurate overstructure, which is analogous to the present FeTe/CdTe.
\\

\textbf{First-principles calculations}:
We performed first-principles calculations to evaluate the impact of $c$-axis compression on the magnetic properties. The calculations were conducted using the projector augmented wave method implemented in the Vienna \textit{ab initio} simulation package \cite{VASP}, with the generalised-gradient-approximation type exchange-correlation functional proposed by Perdew, Burke, and Ernzerhof \cite{PBE}. In our simulations, the lattice parameters were first fixed to the experimental values of the tetragonal phase of Fe$_{1.11}$Te: $a=b=3.8253$ \AA, $c=6.2787$ \AA, and $z_{\rm Te}=0.28207$ \cite{koz2013low}. We then systematically reduced the $c$-axis lattice parameter by $x$\% ($x=0.5, 1.0, 1.5, 2.0$, and $2.5$) keeping all other structural parameters unchanged. All the calculations were performed within the magnetic unit cell ($2\times1\times1$ relative to the primitive cell), using a plane-wave cutoff energy of 500 eV, and a Monkhorst-Pack $k$-grid of $12\times24\times12$.\\

\textbf{Upper critical field}:
We estimated the upper critical field, $H_{\rm c2}$, of FeTe/CdTe film to get an insight into the superconducting properties. Extended Data Figs. \ref{fig:2DSC}b and \ref{fig:80nm}a show $T$-dependence of $R$ at magnetic field $\mu_0H$ = 14 T for both in-plane and out-of-plane orientations for FeTe40/CdTe and FeTe80/CdTe, respectively. The superconductivity is not completely destroyed by the application of a 14 T magnetic field. Instead, small shifts in the $R$-$T$ curves towards lower temperatures are observed. The $H$-$T$ phase diagrams for both field configurations, constructed from the temperatures at which $R$ reduces to half of the normal-state value, are presented in Extended Data Figs. \ref{fig:2DSC}c and \ref{fig:80nm}b. The anisotropy in $H_{\rm c2}$ appears to be relatively small, similar to those seen in Te-rich FST thin films grown on CdTe substrates \cite{Sato2024Non-Fermi}.

We performed analyses based on the Werthamer-Helfand-Hohenberg (WHH) formula for a type-II superconductor in the dirty limit, incorporating both orbital and Pauli pair-breaking effects \cite{WHH1966}:
\begin{align}
\ln{\dfrac{1}{\bar{t}}} &= \left( \dfrac{1}{2} + \dfrac{i\lambda_{\rm SO}}{4\gamma} \right)\psi\left( \dfrac{1}{2} + \dfrac{\bar{h} + \lambda_{\rm SO}/2 + i\gamma}{2\bar{t}} \right)\nonumber\\
    &+ \left( \dfrac{1}{2} - \dfrac{i\lambda_{\rm SO}}{4\gamma} \right)\psi\left( \dfrac{1}{2} + \dfrac{\bar{h} + \lambda_{\rm SO}/2 - i\gamma}{2\bar{t}} \right) - \psi\left( \dfrac{1}{2}\right).
\end{align}
Here, $\psi(x)$ is the digamma function, $ \bar{t} = T/T_{\rm c,0}$, $T_{\rm c,0}$ is $T_{\rm c}$ under zero field, $\gamma \equiv [(\alpha_{\rm M}\bar{h})^{2} - \lambda_{\rm SO}^2]^{1/2}$, and $\lambda_{\rm SO}$ parameterises the strength of spin-orbit scattering. Additionally, 
\begin{equation}
    \bar{h} = \dfrac{4}{\pi^2}\dfrac{H_{\rm c2}}{(-dH_{\rm c2}/dt)_{t=1}}
\end{equation}
was estimated from linear fittings near $\bar{t} = 1$. For simplicity, we fixed $\lambda_{\rm SO} = 1$, which was also employed in FST thin films \cite{Sato2024Non-Fermi}. Superconductivity is predominantly limited by the Pauli effect, evidenced by the extracted large Maki parameter exceeding unity, $\alpha_{\rm M}$ = $\sqrt{2}H_{\rm c2}^{\rm orb.}/H_{\rm c2}^{\rm P} = 11$, where $H_{\rm c2}^{\rm orb.}$ and $H_{\rm c2}^{\rm P}$ are critical fields limited by the orbital and Pauli effects, respectively. This dominant role of the Pauli effect was also reported in FST thin films with high Te content \cite{Sato2024Non-Fermi}, suggesting a smooth evolution of superconducting properties near the FeTe end.\\

\textbf{ACKNOWLEDGMENTS}\\
We thank Xiuzhen Yu, Yingming Xie, Masataka Mogi, and Lixuan Tai for discussions. This work was supported by JSPS KAKENHI (Grant Numbers 24K17020, 22K18965, 23H04017, 23H05431, 23H05462, 24H00417, 24H01652, and 25H02126.), JST FOREST (Grant Number JPMJFR2038), JST CREST (Grant Numbers JPMJCR1874 and JPMJCR23O3), Mitsubishi Foundation, Sumitomo Foundation, Tanaka Kikinzoku Memorial Foundation, and the RIKEN TRIP initiative (Many-body Electron Systems). This work was performed under the approval of the Photon Factory Program Advisory Committee (Proposal No. 2022G551).\\

\textbf{AUTHOR CONTRIBUTIONS}\\
Y.T. conceived and supervised the project. Y.S. and S.N. synthesised samples with help from I.B., R.Y., M.Kawamura, A.T., N.K., and M.Kawasaki. Y.S. and S.N. performed transport and atomic force microscopy experiments. Y.S. conducted STEM experiments with MST. Y.S., S.N., S.K., and H.S. conducted x-ray diffraction experiments under the guidance of T.A. T.N. and R.A. performed the first-principles calculations. Y.S. analysed the data and all authors contributed to the interpretation. Y.S. wrote the manuscript with input from all authors.\\

\textbf{COMPETING INTERESTS}\\
The authors declare no competing interests.

\newpage
\begin{center}
\begin{figure*}[tb]
\includegraphics[scale=0.72]{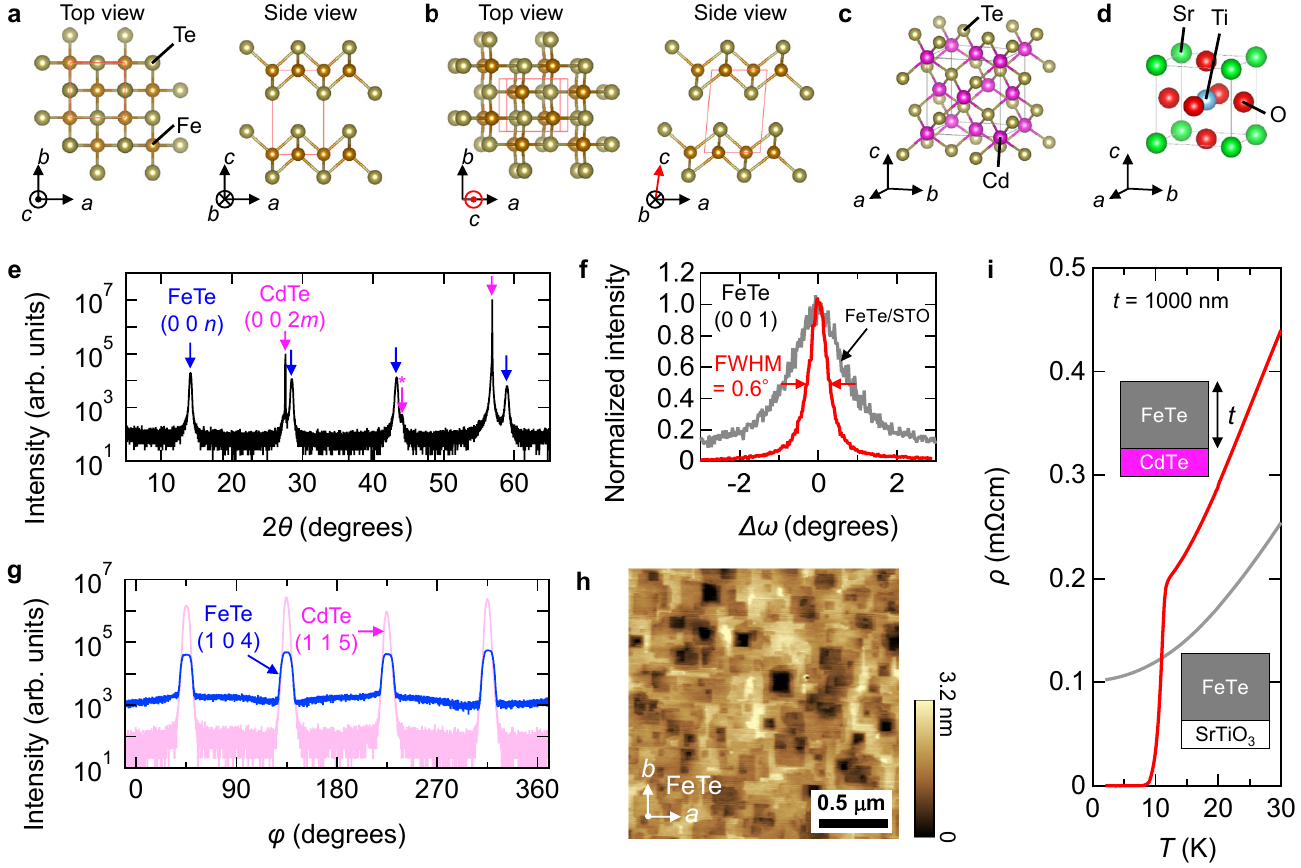}
\caption{\label{fig:Xtal} \textbf{Crystal structures, characterisation, and substrate-selective superconductivity of FeTe films.} \textbf{a}, High-temperature tetragonal and \textbf{b}, low-temperature monoclinic crystal structures of FeTe \cite{VESTA}. The red boxes indicate the unit cell. The in-plane anisotropy $a/b$ and the tilt angle $\pi/2 - \beta$ are slightly exaggerated to emphasize the monoclinic distortion. \textbf{c}, Crystal structure of CdTe. \textbf{d}, Crystal structure of SrTiO$_3$ (STO). \textbf{e}, X-ray diffraction (XRD) $\theta$-2$\theta$ profile of for a FeTe/CdTe film. The blue and magenta arrows indicate FeTe (0 0 $n$) and CdTe (0 0 2$m$) reflections, respectively, where $n$ and $m$ are integers. The asterisk demotes an impurity phase of NaCl-type CdTe. \textbf{f}, XRD rocking curves for the FeTe (0 0 1) reflection for FeTe/CdTe (red) and FeTe/STO (grey) films. \textbf{g}, XRD azimuthal profiles for the asymmetric Bragg points of FeTe (1 0 4) (blue) and CdTe (1 1 5) (magenta). \textbf{h}, Atomic force microscopy image for an FeTe/CdTe film after vacuum annealing. The scale bar represents 0.5 $\mu$m. \textbf{i}, Temperature dependence of resistivity for FeTe films of 1000 nm thickness grown on CdTe and STO.}
\end{figure*}
\newpage
\begin{figure*}[tb]
\includegraphics[scale=0.73]{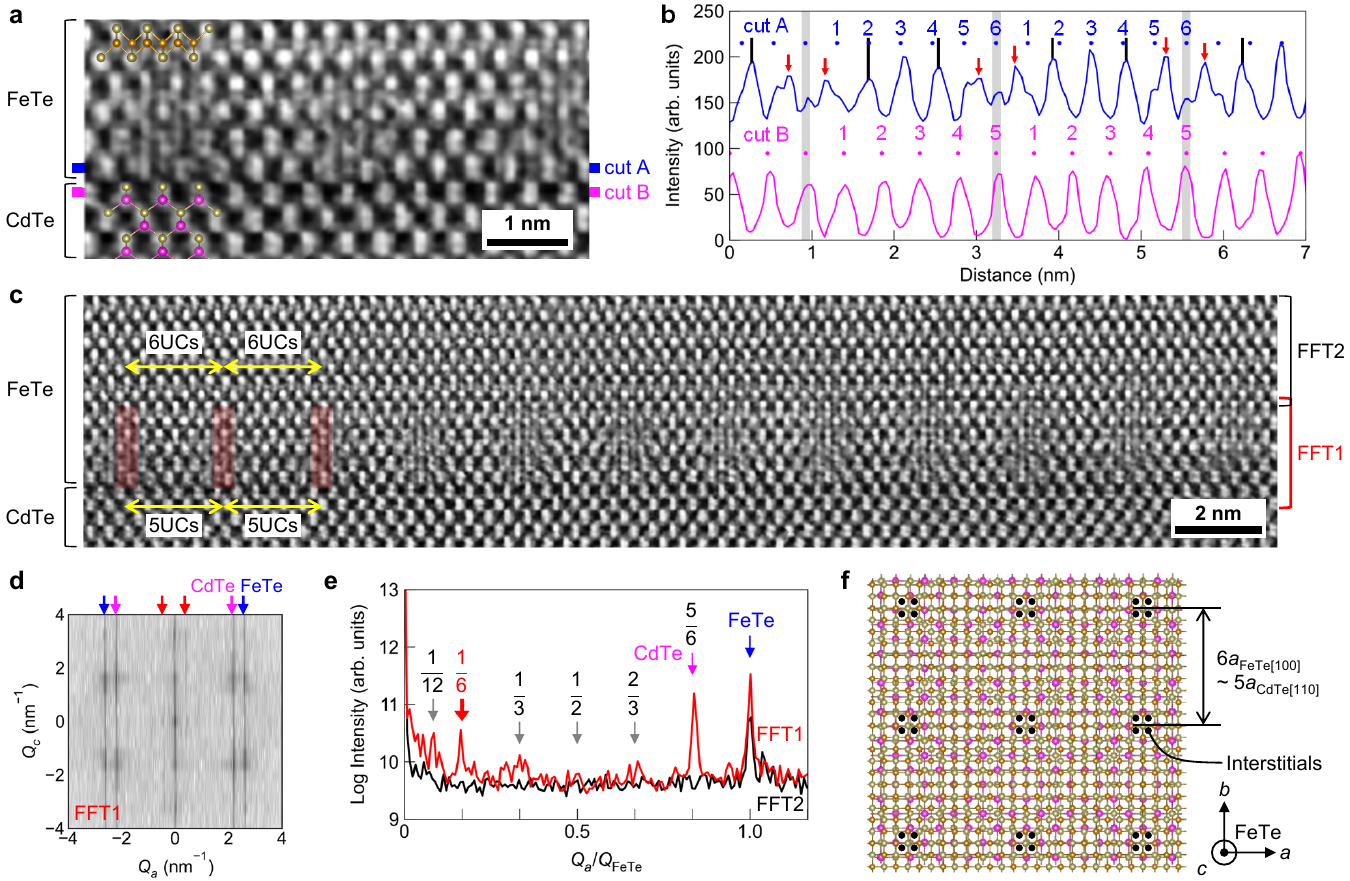}
\caption{\label{fig:STEM} 
}
\end{figure*}
\end{center}
\vspace*{-40pt}
\textbf{FIG. 2. Real-space observation of self-organised periodic interstitial modulation near FeTe/CdTe interface.} \textbf{a}, Cross-sectional HAADF-STEM image near an FeTe/CdTe interface. The yellow, brown, and magenta spheres indicate Te, Fe, and Cd atoms, respectively. \textbf{b}, Integrated-intensity profiles along the cut A and B indicated in \textbf{a}. The dots above the profile indicate the equilibrium position for each atom. The grey areas indicate the sites where the interfacial ions approach most closely. The red arrows represent the interfacial interstitials. The vertical solid black lines represent peak position for the sites 2 and 4, showing a slight modulation from the equilibrium towards the nearest sites. \textbf{c}, Wide area HAADF-STEM image near an FeTe/CdTe interface. The red shaded region indicates the position of the periodic interstitials. The yellow arrows denote the period of modulation, which corresponds to 6 unit cells (UCs) of FeTe, which approximately equals to 5 UCs of CdTe. Areas labelled as FFT1 and FFT2 correspond to the regions where fast Fourier transformation (FFT) was performed. \textbf{d}, FFT image taken within FFT1 in \textbf{c}. The blue, magenta, and red arrows indicate Bragg peaks for FeTe, CdTe, and the interfacial modulation, respectively. \textbf{e}, FFT profiles integrated along $Q_c$ as a function of $Q_a$ divided by the wave number $Q_{\rm {FeTe}}$, corresponding to a Bragg peak of FeTe. The red and black lines correspond to the results taken in FFT1 and FFT2, respectively. The blue, magenta, and red arrows are the same as \textbf{d}, while the grey arrows indicate harmonics of the interfacial modulation, which appear at fractions of $Q_{\rm {FeTe}}$. \textbf{f}, Top view schematics of the 6 $\times$ 6 superlattice of the interfacial interstitials.
\newpage
\begin{figure*}[tb]
\includegraphics[scale=0.6]{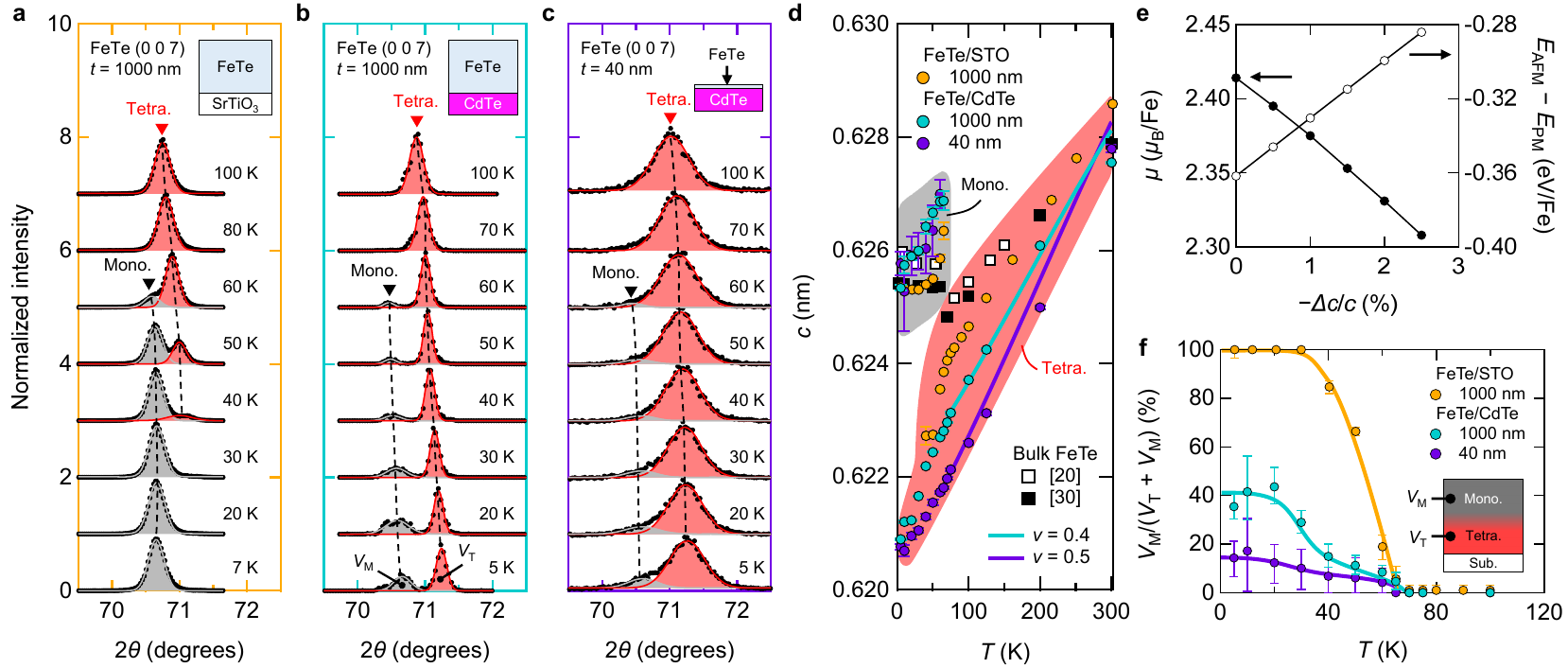}
\caption{\label{fig:FTCT} \textbf{Suppression of monoclinic lattice distortion in FeTe/CdTe films.} \textbf{a-c}, XRD $\theta$-2$\theta$ profiles around the FeTe (0 0 7) Bragg reflection for \textbf{a} FeTe1000/STO, \textbf{b} FeTe1000/CdTe, and \textbf{c} FeTe40/CdTe. Each profile is vertically shifted for clarity. Grey and red curves are Gaussian fittings for the tetragonal and monoclinic phases, respectively. Vertical dashed lines are guides for eyes. \textbf{d}, Temperature dependence of the lattice parameter $c$ deduced from the FeTe (0 0 7) reflections. Orange, cyan, and purple markers denote FeTe1000/STO, FeTe1000/CdTe, and FeTe40/CdTe, respectively. White and black squares are $c$ for bulk single crystals reported by neutron \cite{neutron2009bao} and XRD \cite{XRD2011Xiao} studies. Cyan and purple lines are simulations for the Poisson ratios $\nu$ = 0.4 and 0.5, respectively, assuming perfect clamping to the CdTe substrate. The data highlighted in red and grey areas are those for tetragonal and monoclinic phases, respectively. The error bars represent the standard deviation derived from fittings. \textbf{e},  Fe moments $\mu$ and energy difference between AFM and paramagnetic phases, $E_{\rm AFM} - E_{\rm PM}$, as a function of out-of-plane compressive strain $-\Delta c/c$, obtained by first-principles calculation. \textbf{f}, Temperature dependence of the volume fraction of the monoclinic phase, $V_{\rm M}/(V_{\rm T} + V_{\rm M})$. The inset schematically depicts a film with a phase separation into monoclinic and tetragonal. The lines are guides for the eyes. The error bars represent the standard deviation derived from fittings.}
\end{figure*}

\begin{figure*}[tb]
\includegraphics[scale=0.75]{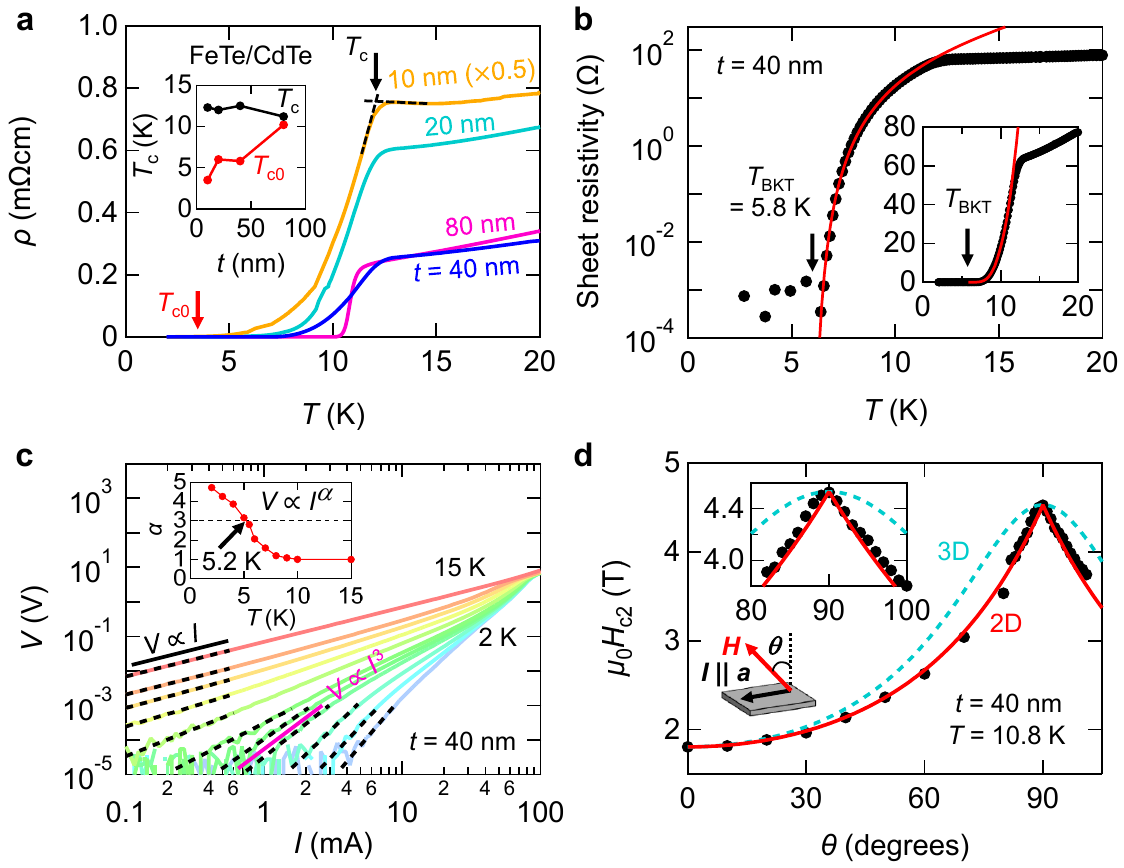}
\caption{\label{fig:SC}
\textbf{Two-dimensional superconductivity in FeTe/CdTe films.} \textbf{a}, Temperature dependence of resistivity with different thickness $t$. The black and red arrows indicate the onset and zero-resistivity critical temperatures, $T_{\rm c}$ and $T_{\rm c0}$, respectively, for the $t$ = 10 nm sample. The inset shows $t$ dependence of $T_{\rm c}$ and $T_{\rm c0}$. \textbf{b}, Temperature dependence of sheet resistivity of the $t$ = 40 nm thin film on logarithmic (main panel) and linear (inset) scales at zero field. The red curve represents fitting result using Halperin-Nelson (HN) formula with $T_{\rm BKT}$ = 5.8 K. \textbf{c}, $I$-$V$ curves in the log-log scale taken with 1-ms-pulsed currents. The measurements were performed at $T$ = 2-10 K in 1 K increments, and additionally at 5.5 K and 15 K. The dashed lines represent linear fittings. The solid black and magenta lines show slopes for $V \propto I$ and $V \propto I^3$, respectively. The inset shows temperature dependence of $\alpha$. The exponent crosses $\alpha$ = 3 at around $T_{\rm BKT}$ = 5.8 K. \textbf{d}, Polar angle $\theta$ dependence of $H_{\rm c2}$. The red solid and cyan dashed lines represent the theoretical 2D Tinkham and 3D anisotropic mass models, respectively. The inset panel shows an expanded view around $\theta$ = 90$^{\circ}$. The inset cartoon illustrates the experimental configuration.}
\end{figure*}

\renewcommand{\figurename}{Extended Data Fig.}
\setcounter{figure}{0}

\begin{figure*}[tb]
\includegraphics[scale=0.85]{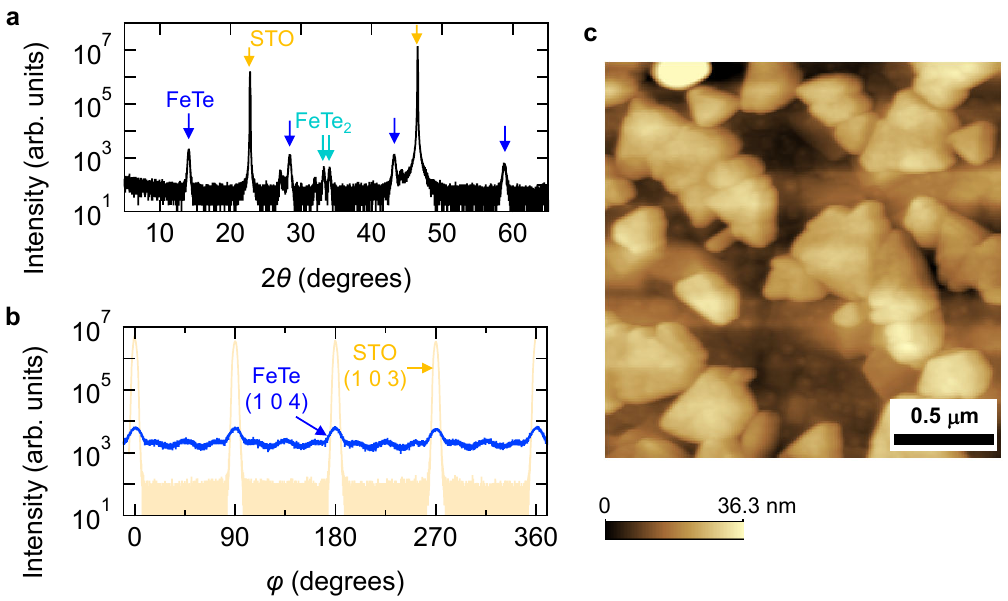}
\caption{\label{fig:FTSTO} \textbf{Characterisation of an FeTe/STO film.} \textbf{a}, XRD $\theta$-2$\theta$ profile. Blue and orange arrows indicate Bragg peaks of FeTe and STO, respectively. Cyan arrows indicate peaks of FeTe$_2$ capping layer. \textbf{b}, XRD azimuthal profiles of asymmetric Bragg reflections FeTe (1 0 4) (blue) and STO (1 0 3) (orange). \textbf{c}, Atomic force microscopy image of an as-grown FeTe/STO film. The scale bar represents 0.5 $\mu$m.}
\end{figure*}

\begin{figure*}[tb]
\includegraphics[scale=1]{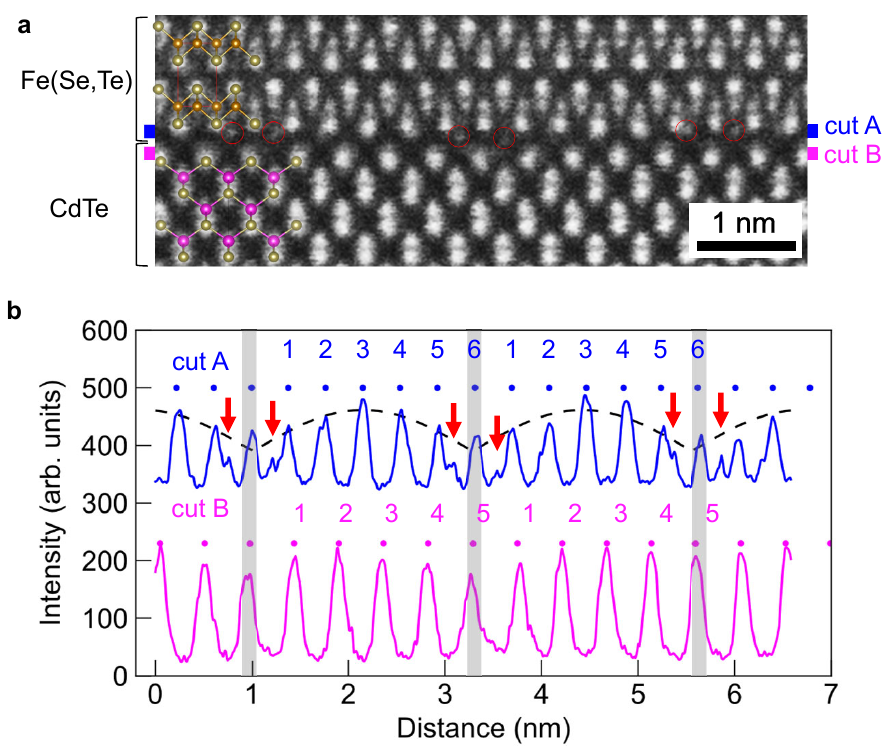}
\caption{\label{fig:STEM_FST} \textbf{Real-space observation of periodic interstitial modulation near FeSe$_{0.1}$Te$_{0.9}$/CdTe interface.} \textbf{a}, Cross-sectional HAADF-STEM image near an FeSe$_{0.1}$Te$_{0.9}$/CdTe interface. Yellow, brown, and magenta spheres indicate Te/Se, Fe, and Cd atoms, respectively. Red circles represent interfacial interstitials. \textbf{b}, Integrated-intensity profiles along the cut A and B, as indicated in \textbf{a}. The dots above the profiles indicate the equilibrium position for each atom. Grey areas indicate the sites where the interfacial ions approach most closely. Red arrows represent interfacial interstitials. The dashed line denotes the amplitude modulation extracted by a sinusoidal fitting with a wave length of $6a_{\rm FST[100]}$.}
\end{figure*}

\begin{figure*}[tb]
\includegraphics[scale=0.62]{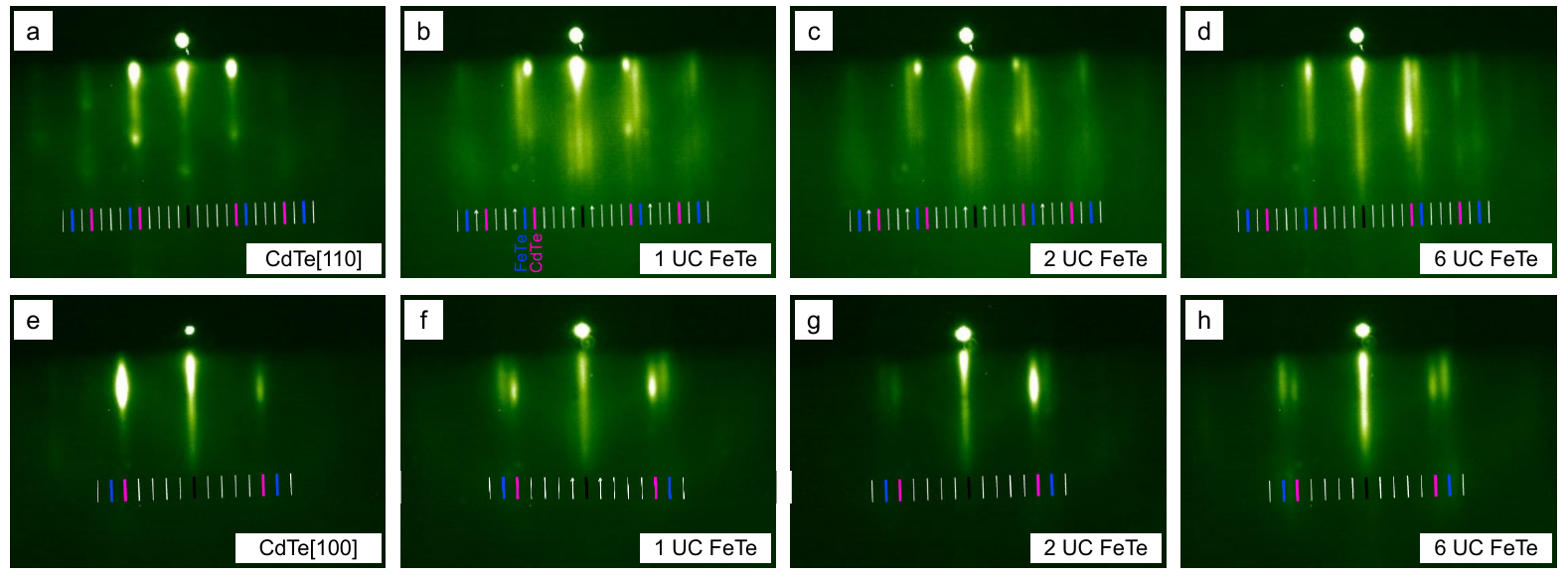}
\caption{\label{fig:RHEED} \textbf{Higher-order epitaxy observed by RHEED.} RHEED images taken with the electron beam parallel to \textbf{a}-\textbf{d}, CdTe[110] $||$ FeTe[100] and \textbf{e}-\textbf{h}, CdTe[100] $||$ FeTe[110]. The images correspond to the CdTe substrate (\textbf{a} and \textbf{e}), 1 UCs FeTe (\textbf{b} and \textbf{f}), 2 UCs FeTe (\textbf{c} and \textbf{g}), and 6 UCs FeTe (\textbf{d} and \textbf{h}). The bottom bars equally spaced are guide for eyes, where the magenta and blue bars denote the streak position of CdTe and FeTe, respectively. The white arrows denote the additional streaks, corresponding to the $6\times6$ modulation. The film thicknesses are estimated based on the growth rate and deposition time.}
\end{figure*}

\begin{figure*}[tb]
\includegraphics[scale=0.87]{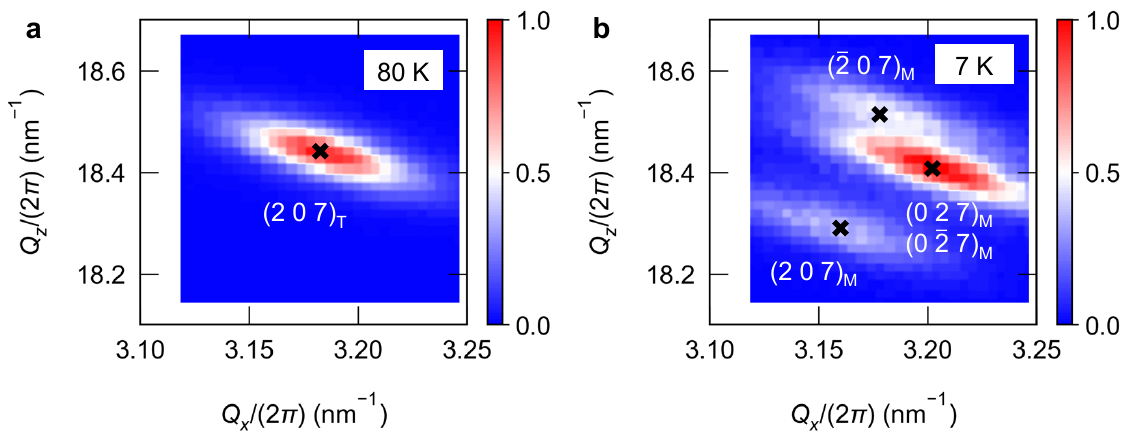}
\caption{\label{fig:RSM} \textbf{Tetragonal to monoclinic structural transition in an FeTe/SrTiO$_3$ film.} \textbf{a, b}, Reciprocal space mappings around the FeTe (2 0 7) Bragg reflection at 80 K and 7 K, respectively. The subscripts T and M indicate the reflection indices in the tetragonal and monoclinic phases, respectively.}
\end{figure*}

\begin{figure*}[tb]
\includegraphics[scale=0.7]{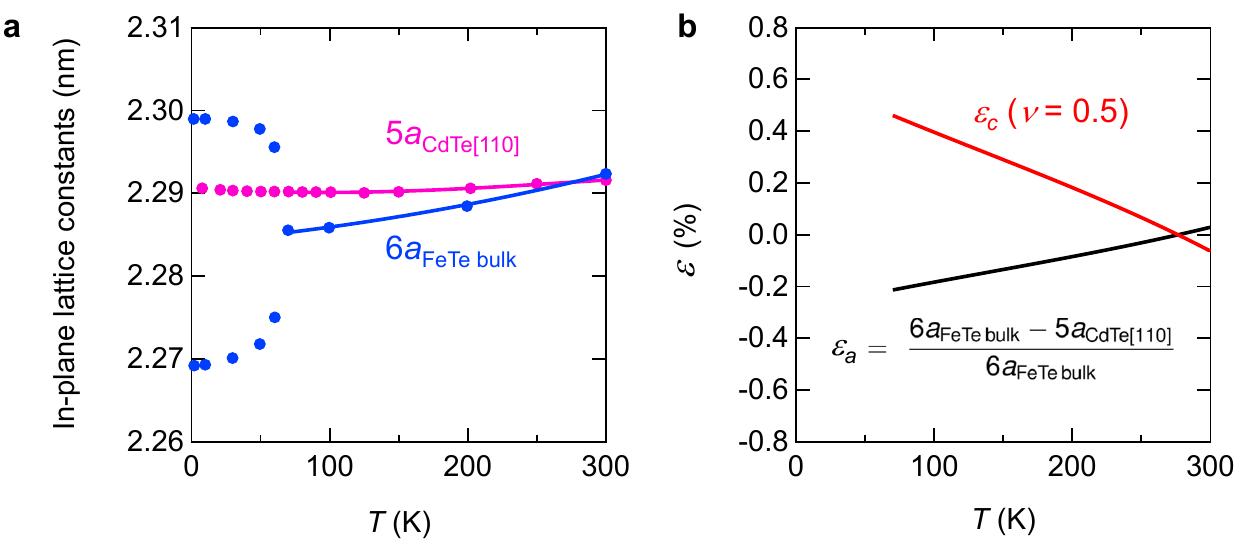}
\caption{\label{fig:thermal} \textbf{Strain induced in FeTe films due to the mismatch in thermal expansion from CdTe substrate.} \textbf{a}, Temperature dependence of the in-plane lattice parameters of bulk FeTe, $a_{\mathrm {FeTe\hspace{0.5mm}bulk}}$ \cite{XRD2011Xiao} and CdTe, $a_{\mathrm {CdTe[110]}}$ multiplied by 6 and 5, respectively. $a_{\mathrm {FeTe\hspace{0.5mm}bulk}}$ is normalised by an experimentally determined value for MBE-grown FeTe films at room temperature. $a_{\mathrm {FeTe\hspace{0.5mm}bulk}}$ splits in the monoclinic phase. \textbf{b}, Temperature dependence of the in-plane $\varepsilon_a$ and out-of-plane $\varepsilon_c$ strains induced in FeTe films. $\varepsilon_c$ is calculated assuming $\nu$ = 0.5.}
\end{figure*}

\begin{figure*}[tb]
\includegraphics[scale=0.8]{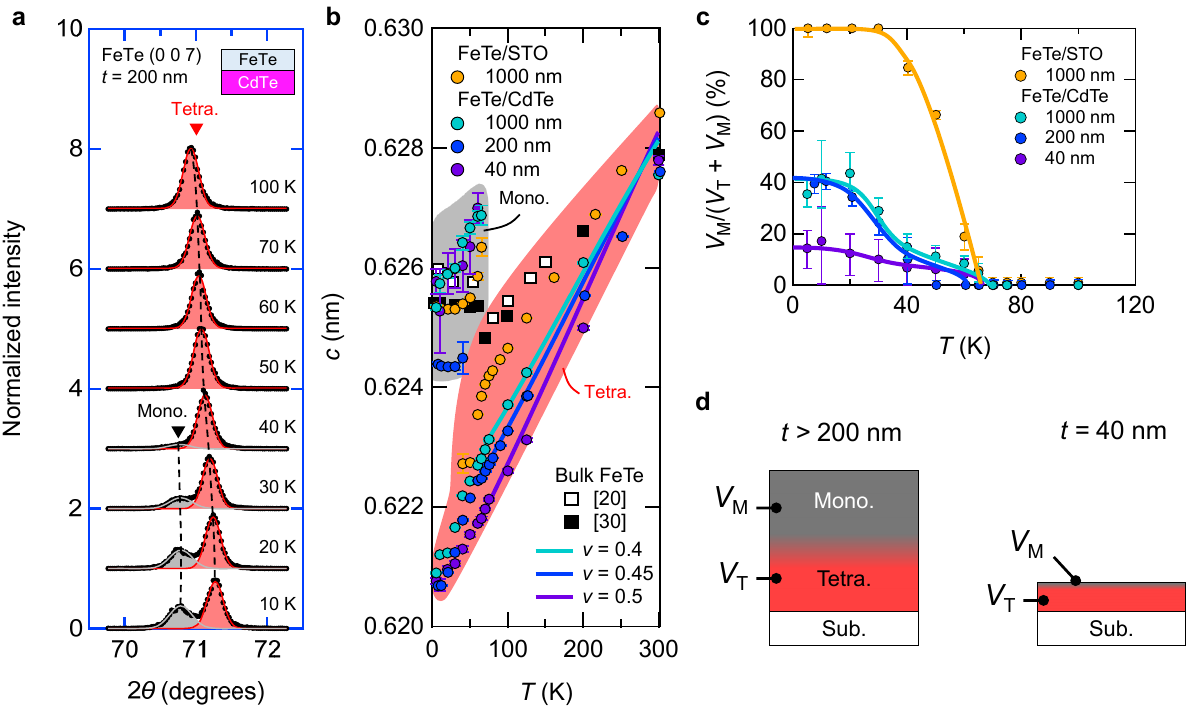}
\caption{\label{fig:200nm} \textbf{Suppression of monoclinic lattice distortion in FeTe200/CdTe sample.} \textbf{a}, XRD $\theta$-2$\theta$ profiles around the FeTe (0 0 7) Bragg reflection for FeTe200/CdTe. Each profile is vertically shifted for clarity. Grey and red curves are Gaussian fittings for the tetragonal and monoclinic phases, respectively. Vertical dashed lines are guides for eyes. \textbf{b}, Temperature dependence of the lattice parameter $c$ deduced from the FeTe (0 0 7) reflections. Orange, cyan, blue, and purple markers denote FeTe1000/STO, FeTe1000/CdTe, FeTe200/CdTe, and FeTe40/CdTe, respectively. White and black squares are $c$ for bulk single crystals reported by neutron \cite{neutron2009bao} and XRD \cite{XRD2011Xiao} studies. Cyan, blue, and purple lines are simulations for the Poisson ratios $\nu$ = 0.4, 0.45, and 0.5, respectively, assuming perfect clamping to the CdTe substrate. The data highlighted in red and grey areas are those for tetragonal and monoclinic phases, respectively. The error bars represent the standard deviation derived from fittings. \textbf{c}, Temperature dependence of the volume fraction of the monoclinic phase, $V_{\rm M}/(V_{\rm T} + V_{\rm M})$. The lines are guides for the eyes. The error bars represent the standard deviation derived from fittings. \textbf{d}, Schematics of FeTe/CdTe films with a phase separation into monoclinic and tetragonal. The monoclinic phase is considerably suppressed in FeTe40/CdTe.}
\end{figure*}

\begin{figure*}[tb]
\includegraphics[scale=0.58]{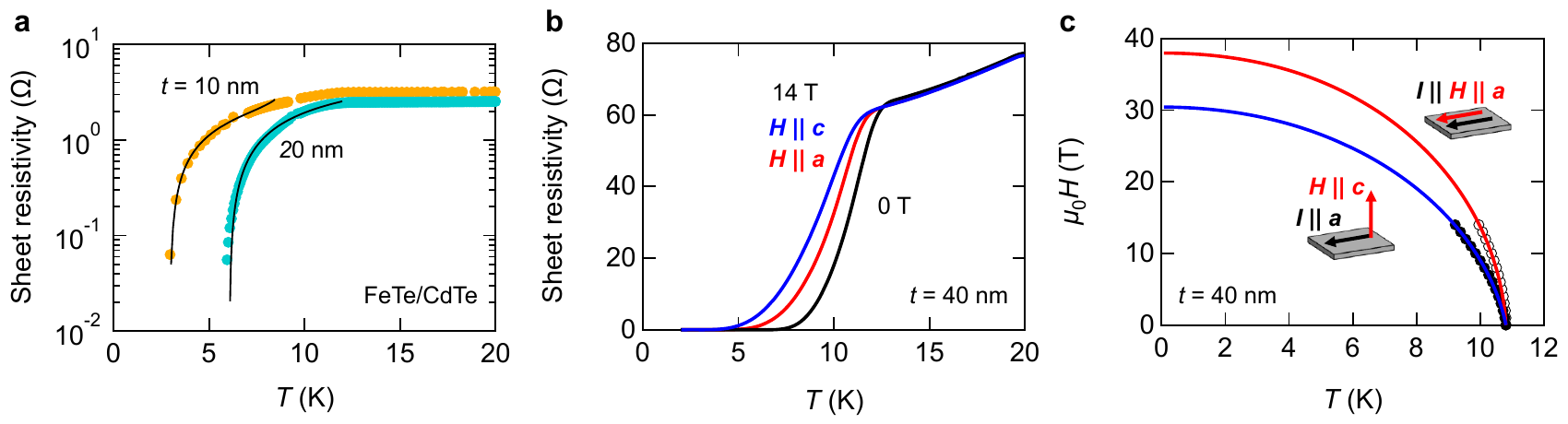}
\caption{\label{fig:2DSC} \textbf{Superconducting properties of FeTe/CdTe thin films.} \textbf{a}, Temperature dependence of sheet resistivity on logarithmic scale at zero field for $t$ = 10 and 20 nm. The black curves represent fitting result using Halperin-Nelson (HN) formula. \textbf{b}, Temperature dependence of sheet resistivity for $t$ = 40 nm at zero field (black), in-plane 14 T field (red), and out-of-plane 14 T field (blue). \textbf{c}, Temperature dependence of the upper critical field for $t$ = 40 nm, defined as the field where the sheet resistivity becomes half of the normal state. Lines show the Werthamer-Helfand-Hohenberg (WHH) fitting results.}
\end{figure*}

\begin{figure*}[tb]
\includegraphics[scale=0.58]{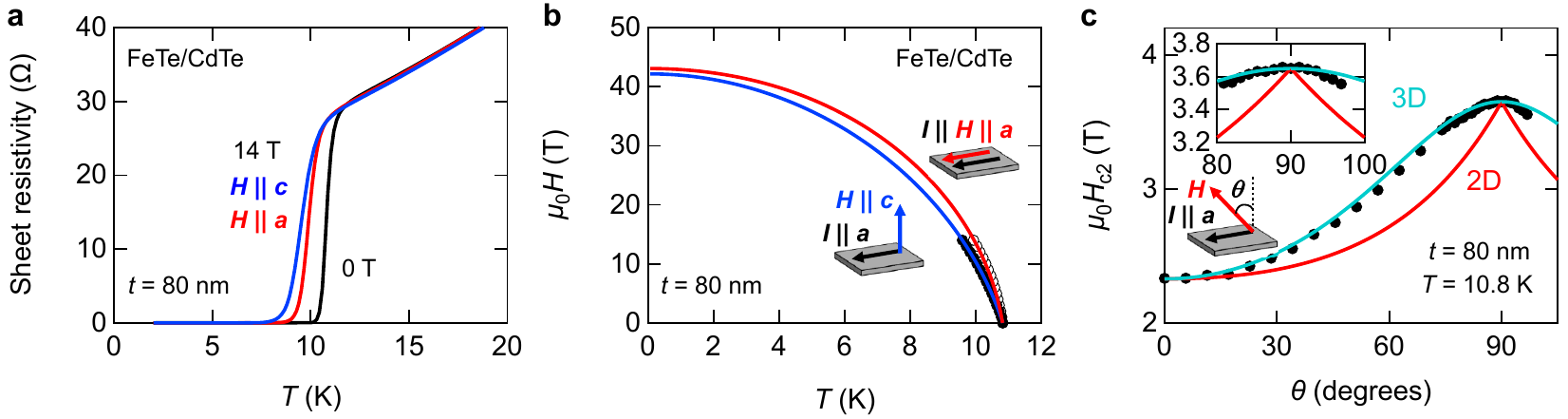}
\caption{\label{fig:80nm} \textbf{Three-dimensional superconductivity in an 80-nm-thick FeTe/CdTe film.} \textbf{a}, Temperature dependence of sheet resistivity for a film with $t$ = 80 nm at zero field (black), in-plane 14 T field (red), and out-of-plane 14 T field (blue). \textbf{b}, Temperature dependence of the upper critical field, defined as the field where the sheet resistivity becomes half of the normal state. Lines show the WHH fitting results. \textbf{c}, Polar angle $\theta$ dependence of $H_{\rm c2}$. The red and cyan lines represent the theoretical 2D Tinkham and 3D anisotropic mass models, respectively. The inset panel shows an expanded view around $\theta$ = 90$^{\circ}$. The inset cartoon illustrates the experimental configuration.}
\end{figure*}

\end{document}